\documentclass[12pt,fleqn]{article}
\usepackage{amsmath}
\usepackage{amsfonts}
\usepackage{amssymb}
\usepackage{graphics,graphicx}
\usepackage[russian]{babel}
\usepackage{cite}
\usepackage{latexsym}

\begin{document}
\title{The problem of small physical parameters and its possible solution}
\author{Sergey G. Rubin \\
National Research Nuclear University "MEPhI", \\ (Moscow Engineering Physics Institute)  \\
sergeirubin@list.ru}

\maketitle

\begin{abstract}
The mechanism of continuous set of different universes formation is elaborated.  It provides  tool to solve the problem of observed smallness of physical parameters. Solution of two puzzles - the hierarchy and the cosmological constant problems based on multidimensional gravity is discussed. 
\end{abstract}

\section{Introduction}
Small physical parameters are often subject of discussions.  Short list of well known parameters is: electron to proton mass ratio, the inflaton mass to the Planck mass ratio, neutrino to electron mass ratio and others. There are many attempts to solve each problem separately. In the paper  \cite{Krause} warped geometry is used to the solution of small cosmological constant problem. The hybrid inflation \cite{LindeHyb} was developed to avoid the smallness of the inflaton mass. The electron to proton mass ratio is discussed in \cite{Trinhammer}. Seesaw mechanism is usually attracted to explain the smallness of neutrino to electron mass ratio \cite{Ibarra}. 

At the same time, universal approach is still absent.  All currently existing models seems to suffer the fine tuning of parameters. Moreover each small parameter needs specific mechanism. 


In this paper a probabilistic mechanism of small parameters formation is developed. It is based on the results of \cite{Gani,RuDef2015}, where deformed extra space with point like defects is studied. As the examples, two issues -  smallness of cosmological constant and hierarchy problem are considered.

The approach is based on small number of well known ideas: extra space, the gravity with higher derivatives and the space-time foam - quantum fluctuation of metric. The latter is already included in the same quantum field theory. Quantum fluctuations of fields and the space geometry in small volumes leads to formation of space domains with evolving content. For an internal observer, this evolution is governed by classical equations of motion.  Widely known and successful model of the chaotic inflation \cite{Linde1} is based on this idea. Effective physical parameters may be different for different final states of the classical evolution \cite{Linde,Suss}. For example, a value of the cosmological constant (CC) depends on the energy density of the inflaton potential. 

Next questions are: how a variety of initial metrics is transformed into a variety of physical parameters specific for different universes? What is the number of final states? The answer depends on initial conditions and a form of field potential. Initial conditions produced by quantum fluctuations represent a continuous set while a number potential minimums is usually supposed to be finite.This means that there exists a finite number of final states and hence a finite number of different universes described by different sets of physical parameters are at our disposal.

Our Universe is described by the specific set of (observed) parameters $\lambda_{obs}$. Some of them are very small. The challenge is to find the set $\{\lambda\}$ and prove that
\begin{equation}\label{in}
\lambda_{obs}\in \{\lambda\}.
\end{equation}
Unfortunately this approach has a weakness. Indeed, even if a number of final states is as huge as $10^{500}$ in the string theory they could be distributed non uniformly in a parameter space and there are no assurance whether \eqref{in} fulfilled. 

The shortcoming mentioned above can be eliminated if we assume that the set of low energy parameters has the cardinality of the continuum. Below a possible way to realize this suggestion is considered.

More definitely, the necessary elements of the approach are:

- space-time foam as the source of any initial metrics. By definition, the space-time foam or quantum foam consists of metric fluctuations which can serve as initial conditions for further  evolution of specific metric \cite{RuInitCond}. This postulate underlines that \textit{all} initial metrics are nucleated.  It seems very reasonable though it can not be proved without a theory of the quantum gravity.

- Extra space does exist. Its dimensionality is not specified here.  An accidental formation of manifolds with various metrics and topologies may be considered as a source of different universes whose variety is connected with a huge number of stationary metrics of an extra space \cite{Brand}, \cite{RuZin}

- The gravity with higher derivatives is a necessary element of a primary theory. The general basis of present study is $f(R)$ gravity. The interest in $f(R)$ theories is motivated by inflationary scenarios starting with the pioneering work of Starobinsky \cite{Star79}. A number of viable $f(R)$ models in 4-dim space that satisfy the observable constraints are proposed in Refs. \cite{Amendola}, \cite{Starob1}, \cite{Odin}. 

If the observable physical parameters (almost) do not  vary with time the same should take place for an extra space metric. Entropic mechanism of the metric stabilization is considered in \cite{RuEntropy}. 
The set of various stationary extra space metrics should have the cardinality of continuum. This set is a final result of the metric evolution governed by the classical equation of motion and hence the final stationary metric could depend on initial configuration. One could keep in mind an analogy with the black holes mass where the metric depends on an initial matter distribution. Each universe as well as ours one is described by specific extra space metric.  

In this paper it is shown that the small parameter problem can be solved on the basis of postulates listed above. To proceed, consider a $n+4$-dim space with an extra $n$-dim metric $G_{AB}$ and initial Lagrangian with primary physical parameters $\alpha_0, \beta_0, ...\lambda_0$. 
After reduction to our 4-dim space we obtain a Lagrangian containing the observable physical parameters $\alpha, \beta,...\lambda$ which depend on the extra metric,  e.g. $\lambda =\lambda (\alpha_0, \beta_0, ...\lambda_0 , G)$, see formulas  \eqref{LLa},  \eqref{ata},\eqref{coefB} as the example.
If the extra metric $G$ is different for different universes, then this secondary parameters varies within an interval,
\begin{equation}
\lambda (\alpha_0, \beta_0, ...\lambda_0, G_-),..., \lambda (\alpha_0, \beta_0, ...\lambda_0, G),..., \lambda(\alpha_0, \beta_0, ...\lambda_0, G_+), \nonumber
\end{equation}
in spite of the primary parameters $\alpha_0, \beta_0, ...\lambda_0$ being fixed.

It points the way to solve the small parameter puzzle. Indeed, if 
$$\lambda (\alpha_0, \beta_0, ...\lambda_0, G_-) <0 \quad \text{and}\quad \lambda(\alpha_0, \beta_0, ...\lambda_0, G_+) >0$$ then there exists a metric $G_0$ such that $\lambda (\alpha_0, \beta_0, ...\lambda_0, G_0)=0$  
provided the set $\{G\}$ of stationary metrics has the cardinality of continuum. Hence there should exist a set of metrics $\{\tilde{G}\}$ such that $\lambda (\alpha_0, \beta_0, ...\lambda_0, \tilde{G})$ is arbitrarily close to zero. Our Universe is connected to some specific extra metric $G_U$. If observed parameter $\lambda_U$ is small then $G_U \in \{\tilde{G}\}$. It will be shown that we are able to find extra space metric $G_U$ of our Universe in that case.
It should be stressed once more that parameter fitting is performed by variation of accidental initial conditions leaving primary parameters fixed. Variation of primary parameters does not solve the problem because the success is usually achieved due to strong fine tuning of this parameters. On the contrary, initial conditions arise as the result of quantum fluctuations so that their variety has natural explanation.

As example, two challenging phenomena - hierarchy of masses and the smallness of the cosmological constant - are considered. A lot of literature is devoted to the solutions for each of them. 
The Hierarchy problem  concerns the smallness of electroweak scale or more definitely the vacuum average of the Higgs field comparing to the Planck scale.
It was studied in various approaches \cite{Dvali}. Up to now each of them can not avoid a fine tuning of primary physical parameters ($\alpha_0, \beta_0, ...\lambda_0$ in our notations). The attempt \cite{Dvali, Randall} to explain the difference between the Planck scale and the electroweak scale is interesting but suffers the same shortage as was shown in \cite{Watanabe}.
In the Randall-Sundrum \cite{Randall} scenarios, 3-branes are embedded in AdS bulk space with a five-dimensional gravity localized on it and negative cosmological constant. As was shown, there is a perfect fine-tuning between brane tension and the cosmological constant. 
The CC problem, the smallness of dark energy widely discussed in literature \cite{Wein}, \cite{Star99}.

\section{Quantum foam and continuous set of different universes}

In this Section the way of continuous set of universes production is discussed. Our Universe is the member of this set with specific (observed) parameters.

From here on, it is assumed that a characteristic scale of extra space is small and its geometry is stabilized shortly after  the Universe creation. The stabilization issue is discussed in   \cite{BRu,KKRu,Green}.

Let us consider a space $M=M_4\times V_n$ with metric
\begin{equation}\label{interval}
ds^2 =\mathfrak{G}_{AB}dZ^A dZ^B = g_{\mu\nu}(x)dx^{\mu}dx^{\mu} + G_{ab}(y)dy^ady^b
\end{equation} 

Here $M_4$ and $V_n$ are the manifolds with metrics $g_{mn}(x)$ and $G_{ab}(x,y)$ respectively.  $x$ and $y$ are the coordinates of the subspaces $M_4$ and $V_n$. We will refer to 4-dim space $M_4$ and $n$-dim compact space $V_n$ as a main space and an extra space respectively. Here the metric has the signature (+ - - - ...), the
Greek indices $\mu, \nu =0,1,2,3$ refer to 4-dimensional coordinates). Latin indices run
over $a,b, ... = 4, 5...$.

Time dependence of the external metric was discussed within the framework of the Kaluza-Klein cosmology and Einstein's gravity \cite{Abbott}. If a gravitational Lagrangian contains terms nonlinear in the Ricci scalar, the extra metric $G_{ab}$ could have asymptotically stationary states \cite{KKRu} 
\begin{equation}\label{statio}
G_{ab}(t,y)\rightarrow G_{ab}(y),
\end{equation}
see also \cite{Carroll}, \cite{Nasri} for discussion.

Consider a gravity with higher order derivatives and action in the form,
\begin{eqnarray}\label{act1}
&& S=\frac{m_D ^{D-2}}{2}\int d^{D}Z \sqrt{|\mathfrak{G}|}\left[f(R) + L_m\right]; \\
&& f(R) = \sum\limits_k {a_k R^k } \nonumber 
\end{eqnarray} 
with arbitrary parameters $a_k,\, k\neq 1$ and $a_1 = 1$ and $D=n+4$. Here $L_m$ is a Lagrangian of matter and $m_D$ is unique scale. Its  lower limit is usually accepted as $\sim 10$TeV.

In the following inequality
\begin{equation}\label{ll}
R_4 \ll R_n
\end{equation}
or more correctly
\begin{equation}\label{lll}
\partial_{\mu}\ll \partial_a
\end{equation}
is assumed. The first suggestion looks natural for the extra space size $L_n < 10^{-18}$ cm as compared to the Schwarzschild radius $L_n \ll r_g \sim 10^6$cm of stellar mass black hole
where the largest curvature in the modern Universe exists.

Using inequality \eqref{ll} the Taylor expansion of $f(R)$ in  Eq. \ref{act1} gives
\begin{eqnarray}\label{act2}
&& S \simeq \frac{m_D ^{D-2}}{2}\int d^{4+n}Z \sqrt{|\mathfrak{G}|}\\ \nonumber 
&&[ R_4(x) f' (R_n) + f(R_n)+L_m]  \nonumber 
\end{eqnarray}
It is assumed that the extra space metric and the fields distribution in the extra space have been evolved to a stationary states that are determined by the stationary classical equations and boundary conditions. The latter depends on initial configuration of the manifold nucleated from the space time foam. It is assumed in the following that a final stationary state is achieved. Its form is governed by the stationary equations which are the subject of study. Besides it is well known that a specific solution to differential equations depends on additional conditions. The latter are just mentioned in this section and will be used in more detail below.

Classical equations for the metric of extra space have the following form
\begin{eqnarray}\label{eqn}
&& R_{ab} f' -\frac{1}{2}f(R)G_{ab} 
- \nabla_a\nabla_b f_R + G_{ab} \square f' =\frac{1}{m_D^{D-2}}T_{ab}, \\
&& \text{+ additional conditions.} \nonumber
\end{eqnarray}
where the first term in \eqref{act2} proportional to $R_4$ is omitted due to inequality \eqref{ll} and $T_{ab}$ denotes the stress tensor of matter. 
Here $\square$ stands by the d'Alembert operator
\begin{equation}
\square =\square_n =\frac{1}{\sqrt{|G|}}\partial_a ( {G}^{ab}\sqrt{|G|}\partial_b),\quad a,b=1,2.
\end{equation}

Up to now we did not specify the number of extra dimensions, its topology, and a form of the function $f(R)$. Now is a proper moment to do it. We limit ourselves by the simplest
choice to support numerically the ideas described above. Let the dimensionality of extra space equal 2, its topology is sphere and the function is
\begin{equation}\label{fR}
f(R)=u_1 (R-R_0)^2
+u_2 .\end{equation}

Evidently, there is a set of solutions to  system \eqref{eqn} depending on additional conditions. Maximally symmetrical extra spaces which are used in great majority of literature represent a small subset of this set. The choice $n=2$ strongly facilitates the analysis. Indeed, if an extra space is 2-dimensional, only one equation in system \eqref{eqn} remains independent. The simplest choice is the equation representing the trace of \eqref{eqn} 
\begin{equation}\label{tr-n}
f'(R_n ) R_n-\frac{n}{2} f(R_n ) +  (n-1)\square_n  f'(R_n ) = T,
\end{equation}
 
The compact 2-dim manifold is supposed to be parameterized by the two spherical angles $y_1=\theta$ and $y_2=\phi$ $(0 \leq\theta \leq \pi, 0 \leq \phi \leq  2\pi)$.
The choice of metric
\begin{equation}\label{metric2}
G_{\theta\theta} = -r(\theta)^2;\quad G_{\phi \phi}= -r(\theta)^2 \sin^2(\theta)
\end{equation}
leads to the Ricci scalar expressed in terms of the extra space radius $r(\theta)$.
As a result explicit form of equation \eqref{tr-n} to be solved numerically is 
\begin{equation}\label{eq2}
 \partial^2_{\theta}R+\cot\theta \partial_{\theta} R =- \frac{1}{2}r(\theta)^2 \left[\left(R_0^2 - R^2 \right) + \frac{u_2 +T}{u_1}\right] .
\end{equation}
As the additional conditions let us fix the metric at the point $\theta =\pi$ 
\begin{equation}\label{bond2}
r(\pi)=r_{\pi} ; \quad r'(\pi)=0 ; \quad  R(\pi)=R_{\pi} ; \quad R'(\pi)=0.
\end{equation} 
From here on we will use the units $m_D = 1$.

Numerical solutions $r_b (\theta)$ for $T=0$ with point like defects and their features are discussed in \cite{Gani}. It was found that due to high nonlinearity of the equation the gravity is able to trap itself in a small region around $\theta =0$ even without matter contribution. 

The metric depends on conditions \eqref{bond2} which are the result of initial conditions. These metrics form a set of cardinality of the continuum. This remark is of extreme importance for the following study.

\section{First Example. The smallness of \\ the cosmological $\Lambda$-term}

The result of previous section is the prove of existence of a continuum set of universes. Next question to be solved is: does our Universe belongs to this set? In particular, we know that it is characterized by the small parameters, some of which are mentioned in the beginning of Introduction. Hence, if this continuum set contains a subset of universes with such small parameters our Universe belongs to this subset. Proof of this statement is a separate task for each small parameter.

Let us apply the former ideas to explain why the cosmological $\Lambda$ term is many orders of magnitude smaller then the Planck scale, $M_{Pl}$.
Consider the Lagrangian of a scalar field
\begin{equation}\label{Lscalar}
L_m =\frac12 \partial_A \varphi \mathfrak{G}^{AB}\partial_B \varphi -U_{\varphi}(\varphi),\quad  U_{\varphi}(\varphi)=\sum_i g_i\varphi^i
\end{equation}
which is contributed to the dark energy.
Equation of motion
\begin{equation}
\square_{4+n}\varphi+U'_{\varphi}(\varphi)=0
\end{equation}
acquires the form
\begin{equation}
\square_{4+n}\tilde{\varphi}+U''_{\varphi}(\varphi_m)\tilde{\varphi}=0,\quad \varphi =\varphi_m +\tilde{\varphi}
\end{equation}
around the potential minimum.
The scalar field is assumed to be almost uniformly distributed in our 4-dim space,
\begin{equation}
\varphi(x,\theta) =\phi (x) Y(\theta) =\varphi_m Y(\theta) +\tilde{\phi}(x)Y(\theta)
\end{equation}
so that the classical equation of motion is
\begin{eqnarray}\label{eqY}
&&Y(\theta)\square_4\tilde{ \phi}(x)+\tilde{ \phi}(x)\square_2 Y(\theta) +Y(\theta) U''_{\varphi}(\varphi_m)\tilde{ \phi}(x)=0. \\
&& \text{+ additional conditions.} \nonumber
\end{eqnarray}
due to \eqref{lll}:
\begin{equation}\label{eqY1}
\square_2 Y(\theta) +U''_{\varphi} (\varphi_m)Y(\theta)=0.
\end{equation}
This equation should be solved together with Eq.\ref{eq2}. The trace of stress tensor has the form 
\begin{equation}
T=G^{ab}T_{ab}=2U_{\varphi}(\varphi)\simeq 2U_{\varphi}(\varphi_m). \end{equation}
Last equality is valid for a 2-dim metric.

Let the functions $G_{ab}(y)$ and $Y(y)$ be the solutions to \eqref{eqn} and \eqref{eqY1} correspondingly. Their substitution into the action leads to the following form of effective action for the proto-Higgs field and gravity
\begin{eqnarray}\label{act3}
 S&=&\pi m_D ^{D-2}\int d^4x d\theta \sqrt{|g(x)G(y)|} R_4(x)f'(R_n(y)) + f(R_n(y)) \\
& +&\pi m_D ^{D-2}\int d^4  x [ K_{\phi}\partial_{\mu}\phi(x)g^{\mu\nu}\partial_{\nu}\phi(x) - U_{\phi} ] , \nonumber \\
&& U_{\phi}(\phi)=\sum_i\tilde{g}_i \phi(x)^i. \nonumber
\end{eqnarray}
where
\begin{eqnarray}\label{LLa}
&& K_{\phi}=\int d\theta \sqrt{|G(\theta)|}Y(\theta)^2 \nonumber
\\
&&\tilde{g}_2=\int d\theta \sqrt{|G(\theta)|}\left[g_1 Y(\theta)^2 -\frac{1}{2}\partial_aY(\theta)G^{ab}\partial_b  Y(\theta)  \right]\\
&&\tilde{g}_{i\neq 2} = g_i \int d\theta \sqrt{|G(\theta)|} Y(\theta)^i  \nonumber
\end{eqnarray}

Comparison of expression \eqref{act3} with the Einstein-Hilbert action
\begin{equation}
S_{EH}=\frac{M^2 _{Pl}}{2} \int d^4x  \sqrt{|g(x)|}(R_4-2\Lambda)
\end{equation}
gives the expression  
\begin{equation}\label{MPl}
M^2 _{Pl}=2\pi m_{D} ^{D-2}\int d\theta \sqrt{|G(\theta)|}f' (R_2 (\theta) )
\end{equation}
for the Planck mass. The term
\begin{equation}\label{density}
\Lambda \equiv \frac{\pi m_D ^{D-2}}{M^2 _{Pl}}\left[ U_{\phi} (v_{\phi}) -\int d\theta \sqrt{|G(\theta)|} f(R_2)\right]
\end{equation}
represents the cosmological $\Lambda$ term. Here $v_{\phi}=v_{\phi}(\tilde {g}_i)$ is a vacuum state of the field $\phi$. The $\Lambda$ term depends on a stationary geometry $G_{ab} (y)$ and hence on the additional conditions.

A scalar field distribution satisfies the classical equation
\begin{equation}\label{eq3}
\cot(\theta)\partial_{\theta}Y(\theta) + \partial^2 _{\theta}Y(\theta)  - U_{\varphi}''(\varphi_m) r (\theta)^2Y(\theta) =0.
\end{equation} 
obtained by substitution metric \eqref{metric2} into equation \eqref{eqY}.
Additional conditions are as follows
\begin{equation}\label{bond4}
Y(\pi)=Y_{\pi} ; \quad Y'(\pi)=0
\end{equation} 
\begin{figure}
\centering
\includegraphics[width=0.7\linewidth]{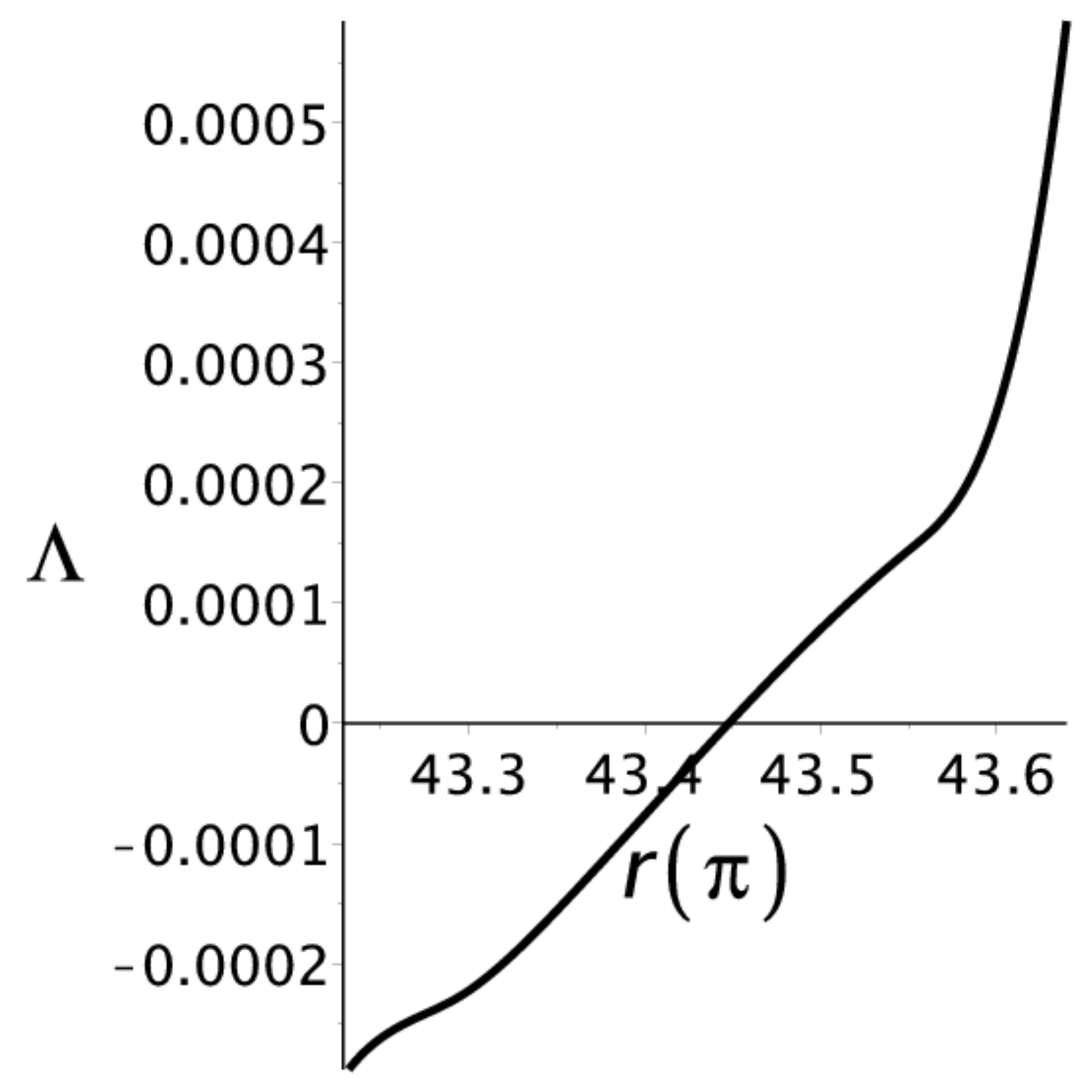}
\caption{The cosmological constant vs. 2-dim extra space radius at point $\theta =\pi$. Here  $u_1=100,u_2=10^{-7},R_0=10^{-3}, g_2=10^{-5},g_3=4.2\cdot 10^{-6}, g_4=3\cdot 10^{-7}$ and $g_i =0$ for $i\neq 2,3,4$. Additional condition is $Y(\pi)=100$. It is assumed that $m_D =1$. $r(\pi)\simeq 43.45$ for our Universe because of the observable smallness of the cosmological constant.}
\label{Lambda}
\end{figure}
As an illustration, numerical solution to Eq.\ref{eq3} is represented in Fig.\ref{Lambda} where specific parameter values are listed.
One can see that the curve intersects zero value at the extra space radius $r(\pi)\simeq 43.45$ which means that 
universes with arbitrary small cosmological constants do form the continuous subset.
Our Universe belongs to this subset. 

According to expression \eqref{MPl}
the D-dim Planck mass is connected to the Planck mass. For the numerical values of the parameters listed in Fig.\ref{Lambda} one can obtain the connection $m_D=M_{Pl}/21.5 $ as the result of numerical calculations.
This allows to fix the radius of the extra space $r(\pi)=43.45/m_D \simeq 900 /M_{Pl}$. The characteristic size of the extra space is large enough to be described classically.

\section{Primary and secondary parameters \\ of a theory}

Previous discussion reveals that some physical parameters acquire their present values as a result of $D$-dim space 1-step reduction to 4-dim space. These parameters are "secondary"\, with respect to primary parameters of initial Lagrangian. This concerns at least the Planck mass, the cosmological Lambda-term and the parameters of the Higgs potential. It would be interesting to study whether parameters of action \eqref{act1} are primary ones indeed. In this section it is shown that a 2-step reduction \cite{RuZin}, $(d+n+4)$-dim space $\rightarrow$ $(n+4)$-dim space $\rightarrow$ 4-dim space, provides additional opportunities.

Another aspect is connected to existence of moderately small parameters like the electron mass to the proton mass ratio, ratio of the inflaton mass to the Planck mass and so on. One can notice that the primary parameters such as $g_i$ in Fig.\ref{Lambda} are also quite small. This may be considered as a defect of the approach developed here. As shown below 2-step reduction can improve the situation significantly.

To proceed, consider a theory with Lagrangian
\begin{eqnarray}\label{act4}
&& S=\frac{m_D ^{D-2}}{2}\int d^{4+n}Z d^d w \sqrt{|\mathfrak{G}_{4+n+d}|}f(R)
\end{eqnarray} 
The manifold $M=M_4 \times V_n \times V_d$ is slightly more complex than previous one. Coordinates $w$ describe new $d$-dim extra space $V_d$ with a metric $G_d (w)$.

Suppose that the Ricci scalar $R_d$ of the extra space $V_d$ satisfies inequality
$$R_d \gg R_{4+n}.$$
Then approximate equality $f(R)\simeq f(R_d (w))$ takes place and classical equations similar to \eqref{eqn} can be obtained by variation action \eqref{act4} with the Lagrangian $f(R_d (w))$. If a metric $G_d (w)$ is a solution to this equations the action is reduced to
\begin{equation}\label{act6}
S=\int d^{4+n}Z\sqrt{\mathfrak{G}_{4+n}}\sum_k \tilde{a}_k R^k _{4+n}(Z)
\end{equation}
\begin{equation}\label{param2}
\tilde{a}_k =\pi m_D^{D-2}\int dw \sqrt{G_d}\frac{f^{(k)}(R_d)}{k!}, \quad d=2
\end{equation}
in $4+n$-dim manifold.
The form of this action is identical to the geometrical part of action \eqref{act1} where the parameters $a_k$ are supposed to be fundamental ones. Essential difference between expressions \eqref{act1} and \eqref{act6} is that parameters \eqref{param2} depend on the geometry of $V_d$ space and one can fit them to desirable values in the same manner as it was done for the observable parameters. Let us demonstrate it on the base of our initial parameters $u_1, u_2$ and $R_0$ which have been considered as primary parameters up to now. 

In the case of quadratic gravity \eqref{fR} with $$f(R)=u_1(R-R_0)^2+u_1 = a_0+a_1 R +a_2 R^2$$ there are three parameters to be determined - $\tilde{a}_0,\tilde{a}_1,\tilde{a}_2$ from formula \eqref{param2}
\begin{eqnarray}\label{ata}
&& \tilde{a}_0 = \int d w\sqrt{G_d}(a_0+a_1 R_d +a_2 R_d^2) \nonumber \\
&& \tilde{a}_1 =\int d w\sqrt{G_d}(a_1 + 2a_2 R_d ^2)\\
&& \tilde{a}_2 =\int d w\sqrt{G_d} a_2 . \nonumber 
\end{eqnarray}
The Lagrangian of the reduced action has the form
\begin{equation}
f=\tilde{a}_0+\tilde{a}_1 R +\tilde{a}_2 R^2 \equiv \tilde{u}_1 (R-\tilde{R}_0)^2 +\tilde{u}_2
\end{equation}
Due to evident connections
\begin{equation}\label{parconnect}
\tilde{u}_1=\tilde{a}_2,\quad \tilde{u}_2=\tilde{a}_0-\frac{\tilde{a}_1 ^2}{4\tilde{a}_2},\quad \tilde{R}_0=-\frac{\tilde{a}_1}{2\tilde{a}_2}
\end{equation}
parameters $u_1 , u_2, R_0$ depend on additional conditions as well.

Average scalar curvature $\langle R\rangle$ of 2-dim space must be much smaller than $m_D^{-2}$ to permit
classical behavior. This is the only important reason to choose the parameter $R_0$ to be rather small. The value $R_0 =10^{-3}$ was postulated in previous sections. If it is a primary physical parameter it may be considered as a defect of the model. But now this parameter depends on
the geometry of extra space and hence on boundary conditions. Its smallness can be substantiated in the same way as discussed in previous sections. More definitely, one should determine those boundary conditions $r_{\pi}$ for which inequality
\begin{equation}\label{R0tilde}
R_0=-\frac{\tilde{a}_1}{2\tilde{a}_2} \ll 1
\end{equation} 
holds.

As a numerical example let us choose specific values for the primary parameters $a_0=0.25, a_1 = -1.2, a_2 =0.5$. Numerical calculations reveal that the secondary parameter $R_0$ of the reduced space intersects zero if boundary condition varies in the interval \begin{equation}\label{interval3}
r(\pi)=1.406 \div 1.411.
\end{equation}
 Thus, we may choose the varying parameter  $R_0$ in \eqref{fR} arbitrarily small assuming that it is a secondary parameter obtained as a result of the first step of reduction. This result will be used in next section.

\section{Second example. Hierarchy problem}
The lesson of section 3 is as follows: there exists a continuous set of secondary parameters, the $\Lambda$ terms,  and each of them is observed in an appropriate universe. It was proved that this set contains a subset with cosmological constants arbitrarily close to zero.

Another problem is the smallness of the Higgs vacuum average $v$. For our Universe, the ratio $v/M_{Pl}=v_{obs}/M_{Pl}= 246/(1.22\cdot 10^{19})\simeq 2\cdot 10^{-17}$ is small parameter that needs an explanation. The question is whether we are able to solve this problem applying the same idea. To this end, let us use the extra space with interval square \eqref{interval} and time-like coordinates of the extra space \cite{timelike}. The shortage of many-time idea and its solution is discussed in \cite{timelike2,timelike3,timelike4}. In the following we suppose that the idea is correct and substitute the minus signs in \eqref{metric2} by plus signs.
The influence of the Higgs field to the extra space metric is supposed to be negligible. 

Consider a Lagrangian $L_H$ of proto-Higgs field $H$
\begin{eqnarray}\label{LH}
&& L_H =\{ \partial_A H^+  \mathfrak{G}^{AB} \partial_B H-U(H^+ H) \}
\\
&& U= b_2\left(H^+ H\right) + b_4 \left(H^+ H\right)^2. \nonumber
\end{eqnarray}
The only conditions imposed on the primary  parameters are $b_2 >0 , b_4 >0$. 
A constant $U(H=0)$ is assumed to be inserted into the function $f(R)$, \eqref{fR}. 

Equation of motion derived from the Lagrangian is
\begin{eqnarray}\label{eqH}
\square_{4+n}H = b_2H +2b_4 \left(H^+ H\right) H.
\end{eqnarray}
As in previous case let us seek for the solution in the form
\begin{equation}\label{x}
H(Z)=H(x,\theta)=\chi(x)W(\theta)
\end{equation}
where the function $\chi(x)$ relates to the Higgs field and the function $W(\theta)$ represents its extra space part.

The proto-Higgs action acquires the form
\begin{eqnarray}\label{Higgs1}
&& S  \simeq \pi m_D ^{D-2}\int d^4xd\theta \sqrt{|{g}||{G}|}\cdot \nonumber  \\
&& \{W(\theta)^2\partial_{\mu}  \chi(x)^+ g^{\mu\nu} \partial_{\nu} \chi(x)+\chi(x)^+ \chi(x) G^{\theta\theta}(\partial_{\theta}W)^2 \nonumber  \\
&&  -b_2 W(\theta)^2\chi(x)^+ \chi(x)  -b_4 W(\theta)^4 [\chi(x)^+ \chi(x)]^2 \}
\end{eqnarray}
after substitution \eqref{x} into \eqref{LH}.
This expression is true if the metric $\mathfrak{G}^{AB}$ is diagonal matrix as it is in our case. Classical equation for the function $W(\theta)$ obtained from action \eqref{Higgs1} has the following form
\begin{eqnarray}\label{eqW}
&&\square_2 W(\theta) +b_2  W(\theta)= -2b_4\chi^+ (x) \chi (x)W(\theta)^3
\end{eqnarray}
where inequality \eqref{lll} is taken into account. The value $ \chi (x)$ is of the order of observed vacuum average  $ \chi (x)\sim v_{obs}\ll 1$ so that nonlinear term may be omitted that gives the equation
\begin{equation}\label{eq4}
\cot(\theta)\partial_{\theta}W(\theta) + \partial^2 _{\theta}W(\theta)  + 2b_2 r (\theta)^2W(\theta) =0.
\end{equation} 
"Wrong" sign before the last term appears due to the time-like character of the extra space coordinates.

Suppose that a solution $W_s (\theta)$ to  equation \eqref{eqW} is known. Then integrating out extra-coordinates $\theta$ in action with Lagrangian \eqref{Higgs1} we obtain the proto-Higgs part of the action
\begin{eqnarray}\label{connections}
S_H && =\int d^4 x \sqrt{g(x)} \{K\partial_{\mu}  \chi(x)^+ g^{\mu\nu} \partial_{\nu} \chi(x) \nonumber  \\
&& - B_2\chi(x)^+ \chi(x)  - B_4 [\chi(x)^+ \chi(x)]^2 \}
\end{eqnarray}
where
\begin{eqnarray}\label{coefB}
&& K= \pi m_D ^{D-2}\int d\theta \sqrt{|G(\theta)|}W_s(\theta)^2 \nonumber \\
&& B_2= \pi m_D ^{D-2}\int d\theta\sqrt{|G(\theta)|}\left[ b_2 W_s(\theta)^2- G^{\theta\theta}(\partial_{\theta}W_s)^2  \right]\\
&& B_4=  \pi m_D ^{D-2}\int d\theta\sqrt{|G(\theta)|}W_s(\theta)^4.  \nonumber 
\end{eqnarray} 
Final transformation
\begin{equation}\label{h}
h(x)\equiv \sqrt{K}\chi
\end{equation}
leads to known form of the Higgs action
\begin{eqnarray}
&& S_H=\int d^4 x \sqrt{g(x)} \{\partial_{\mu}  h(x)^+ g^{\mu\nu} \partial_{\nu} h(x) + \\
&&\lambda  v  ^2 h(x)^+ h(x) -\lambda  [h(x)^+ h(x)]^2    \} \nonumber
\end{eqnarray}
where
\begin{eqnarray}\label{H0}
\lambda  \equiv \frac{4B_4}{K^2}  \label{lambda},\quad
v ^2 \equiv -\frac{B_2}{K\lambda } .\label{v}
\end{eqnarray}

The Hierarchy problem is reduced to the following:
the right hand sides of expressions \eqref{H0} depend on random additional conditions caused by space-time foam so that they form continuous set of universes $\{ \mathcal{U}\}$ with different secondary parameters $\lambda$ and $v$. Observed value of the Higgs vacuum state ($v_{obs}\sim 10^{-17}$ in the Planck units) is very close to zero. Therefore, if the set $\{v\}$ includes $v=0$ with its vicinity,  our Universe might belong to set  $\{ \mathcal{U}\}$. Next step is devoted to prompt realization of this statement.

Equation \eqref{eq4}
with additional conditions $
W({\pi})=W_{\pi} ; \quad W'({\pi})=0
$
describes the behavior of extra part of the proto-Higgs field. Its solution $W_s (\theta)$ depends on additional condition $W_{\pi}$ and the Lagrangian parameters $R_0 , u_1 , u_2$. Let us use the results of previous section and allow variation of the parameter $R_0$ leaving the other parameters constant. Therefore the values $K, B_2, B_4$ and $v$ are also functions of $R_0$.
\begin{figure}
\centering
\includegraphics[width=1.2\linewidth]{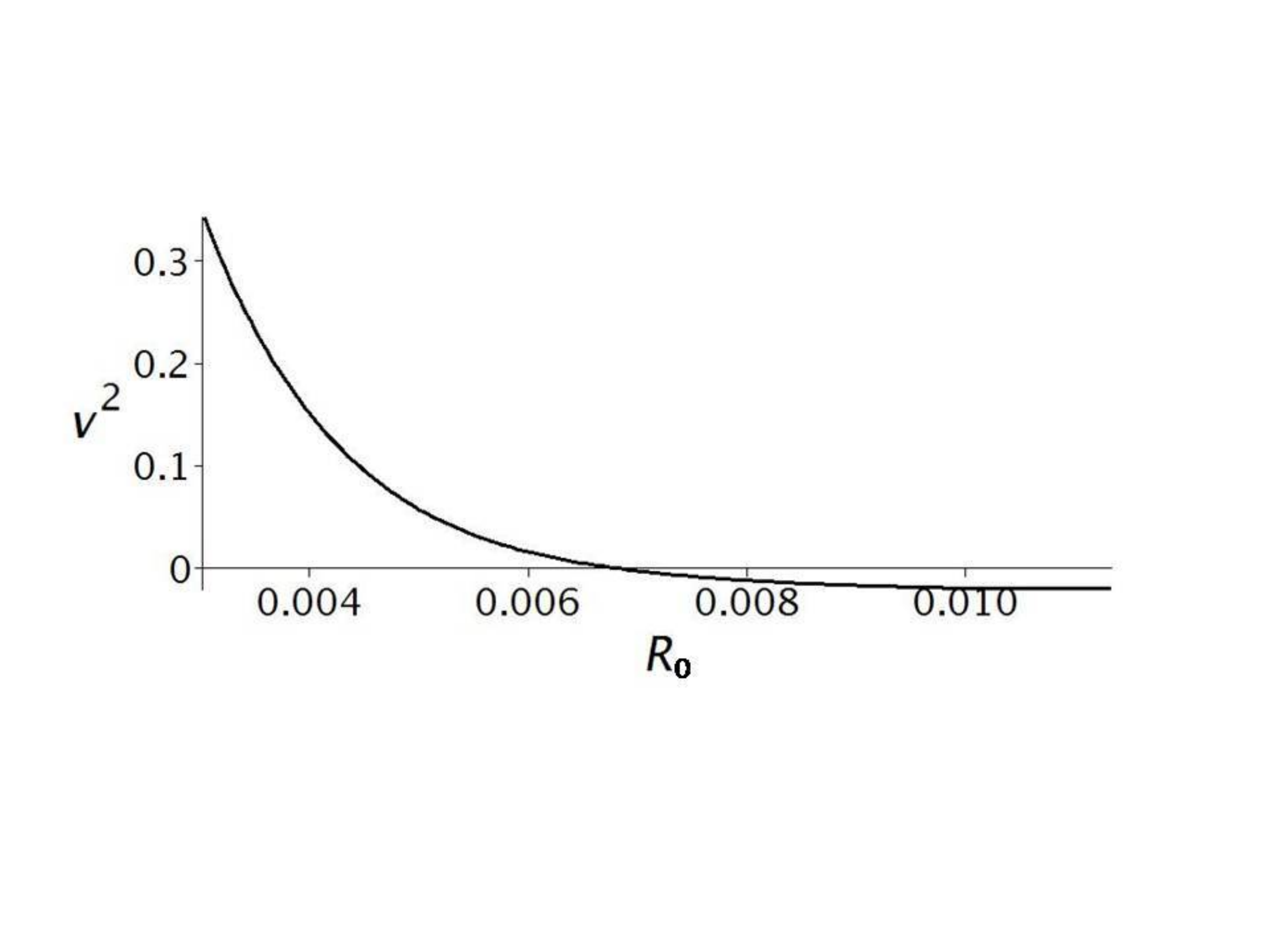}
\caption{The Higgs vacuum average square vs. the parameter $R_0$. $u_1=100,u_2=10^{-7},b_2=10^{-2},b_4=0.1$. }
\label{vSmall}
\end{figure}
Numerical simulations presented in Fig.\ref{vSmall} indicate that the curve intersects zero value at $R_0\simeq 0.068$.

It means that a set of universes with arbitrary small positive Higgs vacuum state do exist. For our Universe $v/M_{Pl}\simeq 2\cdot 10^{-17}$ and hence it belongs to this set.

Numerical results obtained at the end of sections 3,4 and 5 strongly depend on the choice of primary parameter values and hence should be considered as an illustration of the approach. Numerical simulations indicate that main results are mostly insensitive to variation of this  values because it can be compensated by an alternation of additional conditions.

\section{Discussion}

Nowadays, nobody wonders why the properties of our planet are so suitable for life just because enormous variety of different planets is known to exist. 
The same argument can be applied to the physical properties of our Universe. We would not wonder why the observed physical characteristics are so suitable for life if there were a huge variety of different universes.
As was discussed in the Introduction the key issue is that this variety should have the cardinality of the continuum.

The general picture is as follows. Quantum fluctuations of metrics generate universes with different numerical values of physical parameters $ \lambda $. Our universe is one of those universes. 
More precisely, every physical parameter $\lambda $, the electron mass $ m_e $ for example, is a functional $\lambda [G]$ of  extra space metric $G$. Therefore, the set of all metrics $G$ generates a set of values of the electron mass $m_e [G] $. Specific value $m_e = 9,10938292·10^{-31}$kg, observed in our universe, is a representative of the set.

Under this approach, \textit{the justification of the observed parameter value means a solution of the inverse problem of finding a suitable extra space metric}. For example, one should prove that there exists a metric $ G_* $ for which $m_e[G_*]=9,10938292·10^{-31}$kg. In general, the purpose seems difficult to fulfill if a set of metrics is discrete. However, if a set of different metrics has the continuum measure the task is greatly simplified.

For example, if the observed parameter $\lambda_{small}$ is very small it is enough to prove that a set $\{\lambda[G]\}$ includes the value $\lambda[G_*]=0.$ In this case,  if the set has the cardinality of the continuum, then there exists such a metric $G_+$ for which $\lambda[G_+]=\lambda_{small}\lll 1.$

In this paper, the formation mechanism of continuous set of universes with different physical parameters is elaborated. It is also proved that there exists a continuous subset of universes described by small physical parameters such as the cosmological constant and the electroweak scale. Evidently our Universe belongs to this subset. There is no necessity to invent special mechanisms  to explain the smallness of this parameters. 

Models based on the extra space idea are described by primary physical parameters of Lagrangian. Some of them should be fine tuned to solve a problem in question which may be considered as the defect of a model. Meanwhile
2-step reduction considered in section 4 can significantly smoothen the problem.

\section{Acknowledgment}
This work was performed within the framework of the Center FRPP supported by MEPhI Academic Excellence Project (contract № 02.а03.21.0005, 27.08.2013) and was supported by the Ministry of Education and Science of the Russian Federation, Project No.~3.472.2014/K.

\end{document}